\documentclass[journal]{IEEEtran}
\usepackage{cite}
\usepackage{amsmath,amssymb,amsfonts}

\usepackage{algorithmic}
\usepackage{graphicx}
\usepackage{comment}
\usepackage[switch]{lineno} 
\usepackage{textcomp}
\usepackage{multicol}
\usepackage{caption}
\usepackage[table]{xcolor}
\usepackage{tabularx}
\usepackage{siunitx}

\ifCLASSINFOpdf
\else
\fi
\usepackage{amsmath,amssymb,amsfonts}
\usepackage{algorithmic}
\usepackage{graphicx}
\usepackage{comment}
\usepackage[switch]{lineno} 
\usepackage{textcomp}
\usepackage{multicol}
\usepackage{caption}
\usepackage[table]{xcolor}
\usepackage{tabularx}
\usepackage{siunitx}
\begin{document}
%
\title{Near Field Electric (NFE): Energy-efficient, High-speed Communication at Decimeter-range}
%
%
%

\author{Samyadip Sarkar$^{1}$,~\IEEEmembership{Student Member,~IEEE,} Arunashish Datta$^{2}$, Sihun Kim$^{2}$, Amir Mokhtarpour$^{2}$, \\ Shweta Katakdhond$^{2}$,
Shovan Maity$^{2}$, Shreyas Sen$^{1, 2}$,~\IEEEmembership{Senior Member,~IEEE}
\thanks{$^{1}$Elmore Family School of Electrical and Computer Engineering, Purdue University, West Lafayette, IN 47907, USA,  \\ e-mail: \{sarkar46, shreyas\}@purdue.edu}
\thanks{$^{2}$Ixana, 3000 Kent Ave, Suite 1544, West Lafayette, IN 47906, USA 
}
}

\maketitle

\begin{abstract}

Near-field technologies enable contactless payments, building access, automotive keyless entry, and supply chain tracking. Existing approaches face fundamental trade-offs: magnetic-based methods (NFC/NFMI) achieve low power but are limited to sub-megabit rates, while millimeter-wave techniques provide gigabits/sec connectivity at higher power consumption and only centimeter-scale ranges. We demonstrate that near-field electric (NFE) communication breaks this trade-off via capacitive coupling enabled by confined electric fields. NFE simultaneously achieves ultra-low power ($<$1 mW per transceiver), high-speed data throughput ($>$3 Mbps), and configurable decimeter-range (5-30 cm) capabilities previously considered mutually exclusive. Systematic measurements across multiple orientations and configurations show NFE can support decimeter communication coverage. 
The power consumption of 0.4 mW at the transmitter (Tx) and 0.6 mW at the receiver (Rx), when combined is up to $\sim$24$\times$ lower than NFC and $\sim$3$\times$ lower than NFMI while achieving significantly higher data rates, and a couple of orders of magnitude lower power than mm-wave based technique. Testing with symmetrical electrodes across eight orientations validated consistent performance and robustness for practical deployments. Extended-range experiments achieved stable 2 Mbps throughput at 3.5 meters using conductive media, demonstrating NFE's unique ability to leverage environmental conductors. Optimized device design can facilitate achieving an extended range for Body-assisted NFE up to 1 m. Results establish NFE as foundational for next-generation wireless applications where security, low power, and throughput converge, enabling dense IoT deployments, secure payment systems, and high-speed device-to-device communication previously limited by the power-performance trade-off.

\end{abstract}


%
\IEEEpeerreviewmaketitle

\section{Introduction}
The relentless growth of connected devices and the insatiable demand for low-latency, high-throughput links are straining conventional wireless paradigms. Near Field wireless Communication (NFC) enables secure, high-precision communication over very short distances. It is widely used today in contactless payments, access control, and device pairing, while rapidly evolving toward advanced near-field beamforming for 6G applications \cite{zhang20236g, chen20246g, zhao20256g}.
These span from inductive coupling (NFC) to millimeter-wave, each with distinct power-range-rate tradeoffs. These methods include classical Near Field Communication (NFC) (inductive coupling at 13.56 MHz) \cite{liu2024near} for low‑data, secure exchanges and emerging radio‑frequency near‑field techniques that use focused electromagnetic fields from very large antenna arrays to achieve high spatial resolution and capacity. In current practice, NFC remains the dominant consumer-facing near‑field technology—embedded in smartphones, payment cards, and smart tags for contactless payments, ticketing, and secure access—because it is simple, energy‑efficient, and benefits from standardized protocols and a strong ecosystem. Concurrently, researchers are working towards expanding the near‑field concept for wireless communications beyond short, low‑rate links: as antenna arrays grow (e.g., extremely large‑scale arrays envisioned for 6G) \cite{lu2024tutorial}, the classical far‑field plane‑wave assumption breaks down and near‑field channel models, beamfocusing (rather than beamsteering), and new propagation/beamforming algorithms are required to exploit spatially varying phase and amplitude across the array for higher throughput and user multiplexing. Practical deployments today therefore sit on two tracks: standardized NFC for secure short‑range services and experimental near‑field MIMO/beamfocusing prototypes aimed at indoor hotspots, device‑to‑device high‑rate links, and enhanced localization. Key challenges in moving from NFC to high‑capacity near‑field systems include accurate channel estimation in the reactive region, hardware complexity, interference management, and ensuring privacy/security when spatial resolution can reveal precise user locations. Industry practice addresses some of these issues through protocol layering, secure element integration for payments, and conservative power/interaction ranges for consumer devices, while academic and standards communities are actively developing models and measurement techniques for integrating near‑field operation into future cellular standards. Overall, near‑field technologies today combine robust, low‑power NFC deployments with a growing research and prototyping effort to harness near‑field physics for high‑capacity, spatially precise wireless links in next‑generation networks. Existing wireless options trade power for speed and transmission range: magnetic induction (NFC/NFMI) runs very low power but tops out below 1 Mbps, while millimeter‑wave links deliver multi‑gigabit throughput at higher power cost and only over centimeter‑scale distances. Addressing the trade-off between achieving higher data rates at low power consumption yet higher coverage (in the decimeter range) necessitates an immediate  solution.  

This paper introduces Near Field Electric (NFE), a near field communication technique that harnesses capacitive coupling between communication devices, enabled by confined quasistatic electric fields. In this paper, for the first time, we present a framework for NFE technology, and quantifies NFE's performance against state-of-the-art techniques for short-range communication, based on power, range, and data rate.

\section{Background}

\subsection{The Evolution Toward Near-Field Electric Communication}

\textbf{Historical Context:} Near-field wireless communication evolved from RFID (1983) through NFC standardization (2002-2006), achieving widespread adoption in payment systems despite fundamental limitations: 4 cm range and 424 kbps data rates constrained by magnetic induction at 13.56 MHz. While early capacitive-coupling research established the theoretical foundations for electric-field communication, practical implementation remained limited by technological constraints until recently.

\textbf{Technology Convergence Enabling NFE: } Earlier generations of chips and antennas were optimized for magnetic or RF fields, making magnetic field and radio-based communication the standard approaches. The ability to design miniaturized, efficient transceiver chips for electric-field coupling was not practical until recent advances in CMOS processes and packaging. Moreover, NFE requires precise field confinement and noise rejection capabilities that Digital Signal Processing (DSP) and low-power microcontrollers could only handle efficiently in the last decade.

Specifically, NFE emerges now due to critical advances in semiconductor technology and signal processing. Sub-65-nm CMOS processes enable sub-1 mW transceivers achieving multi-Mbps rates—a combination previously impossible for a decade. Simultaneously, modern DSPs can implement adaptive filtering and field confinement at power budgets around 100's of $\mu$W, with low-power microcontrollers now widespread enough to support the sophisticated algorithms required for NFE's precise field control and noise rejection. These converging technological capabilities finally make electric-field-coupling ASICs practical for commercial deployment.


\subsection{Different Near-Field Operating Regions}
The space surrounding a transmitting element is commonly split into three regions that determine how signals behave and how devices should be designed and secured. Closest to the Tx is the reactive near field, where non‑radiative, evanescent fields dominate, and energy is stored rather than propagated; interactions here fall off rapidly with distance and are exploited by tightly coupled near‑field technologies (e.g., inductive or capacitive coupling) because they enable strong, localized links and reduced eavesdropping risk. Beyond that lies the radiative near field or Fresnel region, a transition zone where the field begins to radiate but the wavefront is not yet fully formed into the simple angular pattern seen at long range; in this region, phase and amplitude vary with distance and the size of the Tx, so coupling and beamforming behave differently than in the far field. Farther out is the far field or Fraunhofer region, where the electromagnetic waves settle into stable plane‑wave behavior, and the angular radiation pattern is fixed; here, power density falls off approximately with the square of distance, and conventional antenna‑to‑antenna link equations apply. Approximate boundary formulas can be applied to separate these regions: the Fraunhofer (far‑field) distance is commonly given by
\begin{equation}
    R_F = \frac{2D^2}{\lambda}
\end{equation}
where D is the largest antenna dimension and 
$\lambda$ is the wavelength, while an approximate Fresnel boundary can be expressed as 
\begin{equation}
    R_{Fresnel} \approx 0.62\sqrt{\frac{D^3}{\lambda}}
\end{equation}
Distances smaller than these correspond to near‑field behavior and larger distances to far‑field behavior. These distinctions matter for near‑field communication design and security: reactive near‑field coupling enables highly localized, low‑power, hard‑to‑intercept links, the Fresnel region can allow radiative coupling with complex phase behavior for phased arrays or larger apertures, and the Fraunhofer region is where conventional radiative communications and long‑range eavesdropping become possible. 

\subsection{Near Field Communication (NFC) $\&$ Near Field Magnetic Induction (NFMI)}
Near Field Communication (NFC) uses short‑range electromagnetic waves to exchange data at a few centimeters ($\sim$4 cm). It is widely used for contactless payments, transit, access control, and quick device pairing \cite{liu2024near}. It gained popularity owing to its ubiquitous integration in smartphones and payment ecosystems. NFC chips are now standard in most phones and smartwatches, making mobile wallets and tap‑to‑pay services easy to roll out at scale. Additionally, NFC facilitates user convenience and payment speed by allowing users to tap a phone or card. Furthermore, NFC leverages established payment tokenization and secure element frameworks, giving merchants and banks confidence to adopt it broadly. Low implementation cost across many use cases, as passive NFC tags and reader modules are inexpensive, enabling marketing, access badges, and IoT triggers without a heavy hardware investment. Each of these factors combined to create a strong network effect: more terminals meant more useful phone features, which in turn encouraged more terminals and services. Although it has remained a standard technique,  its limited transmission range ($\sim$4 cm) and supported data rate of $\sim$0.4 Mbps \cite{ST25R3918} make it suboptimal when the system demands higher coverage and lower power during transferring information. 

On the contrary, Near Field Magnetic Induction (NFMI) uses resonant magnetic coils to couple a non‑radiating magnetic field between devices. It is optimized for very short‑range links that are robust near the human body and in metal- and wet-environmental conditions \cite{pal2020nfmi}. Because NFMI relies on magnetic coupling rather than propagating RF waves, it performs well near the human body, through water, and around metal, where conventional RF can fail \cite{li2019survey, thilak2012near, kim2016near}. Moreover, NFMI's non‑radiating nature keeps signals tightly localized, reducing eavesdropping risk and interference with other wireless systems. Although it can support a transmission range extending up to 100-300 cm, its supported data rate of $\sim$0.6 Mbps\cite{NxH2281_product} makes it not an ideal solution when the system demands higher data rates.

\subsection{Millimeter-wave based Techniques}
Among the different techniques for short-range communication, Millimeter-wave (mm-Wave) approaches are rapidly evolving and exploit the reactive and radiative near field for ultra‑high‑speed, spatially focused links, using large arrays, distance-aware beamforming, and robust beam‑management. Mm-Wave-based research has shifted from treating antennas as far‑field radiators to deliberately harnessing near‑field beamforming and the distance domain as additional degrees of freedom, enabling focused energy delivery and finer spatial multiplexing. Contemporary approaches center on extremely large‑scale MIMO (XL‑MIMO)\cite{wang2023extremely, wang2024tutorial} and dense antenna arrays that create spherical wavefronts rather than planar ones, which requires new beam design and training methods that jointly estimate angle and distance to align energy in the near field rather than only steering by angle \cite{lu2024tutorial}. On the hardware side, advances in mmWave antenna arrays and compact beam‑steering elements—including phased arrays, lens antennas, and integrated on‑chip radiators—are making practical, wearable, and form‑factor‑constrained near‑field deployments feasible while preserving wide instantaneous bandwidths \cite{zhang2022beam}. Robustness is another active thread: algorithms that account for aperture perturbations, manufacturing tolerances, and near‑field steering vector sensitivity are being developed to maintain beamforming gain under real‑world imperfections, improving reliability in multi-user scenarios. Together, these strands—distance‑aware beamforming, position‑aided training, XL‑MIMO array design, and robust beam management—form the current mmWave near‑field systems. However, they also possess open challenges: balancing training overhead against latency, designing low‑power RF front ends for dense arrays, and creating scalable protocols that exploit near‑field confinement without sacrificing interoperability with legacy far‑field systems. It is worthwhile to note that though mm-wave based techniques has potential to offer high speed connectivity (6.25 Gbps), their limited transmission range ($\sim$1 cm) and higher power consumption ($\sim$ 55 mW in Tx mode, 30 mW in Rx mode, values at 3.125 Gbps)\cite{ST60A2_product} makes them sub-optimal solutions when the reliable operation over extended range ($\sim$ decimeters) with an extended battery life is a primary concern. 

\subsection{Opportunities for Electric-field based Systems}
To address the aforementioned limitations of existing near-field-based communication techniques, Near Field Electric (NFE)-based communication is positioned to bridge the gap between ultra-secure, low-power wireless networks and high-speed data exchange. Its unique combination of low power consumption, security, and throughput makes it a strong candidate for next-gen wireless networks.

\begin{figure}[ht]
\centering
\includegraphics[width=0.48\textwidth]{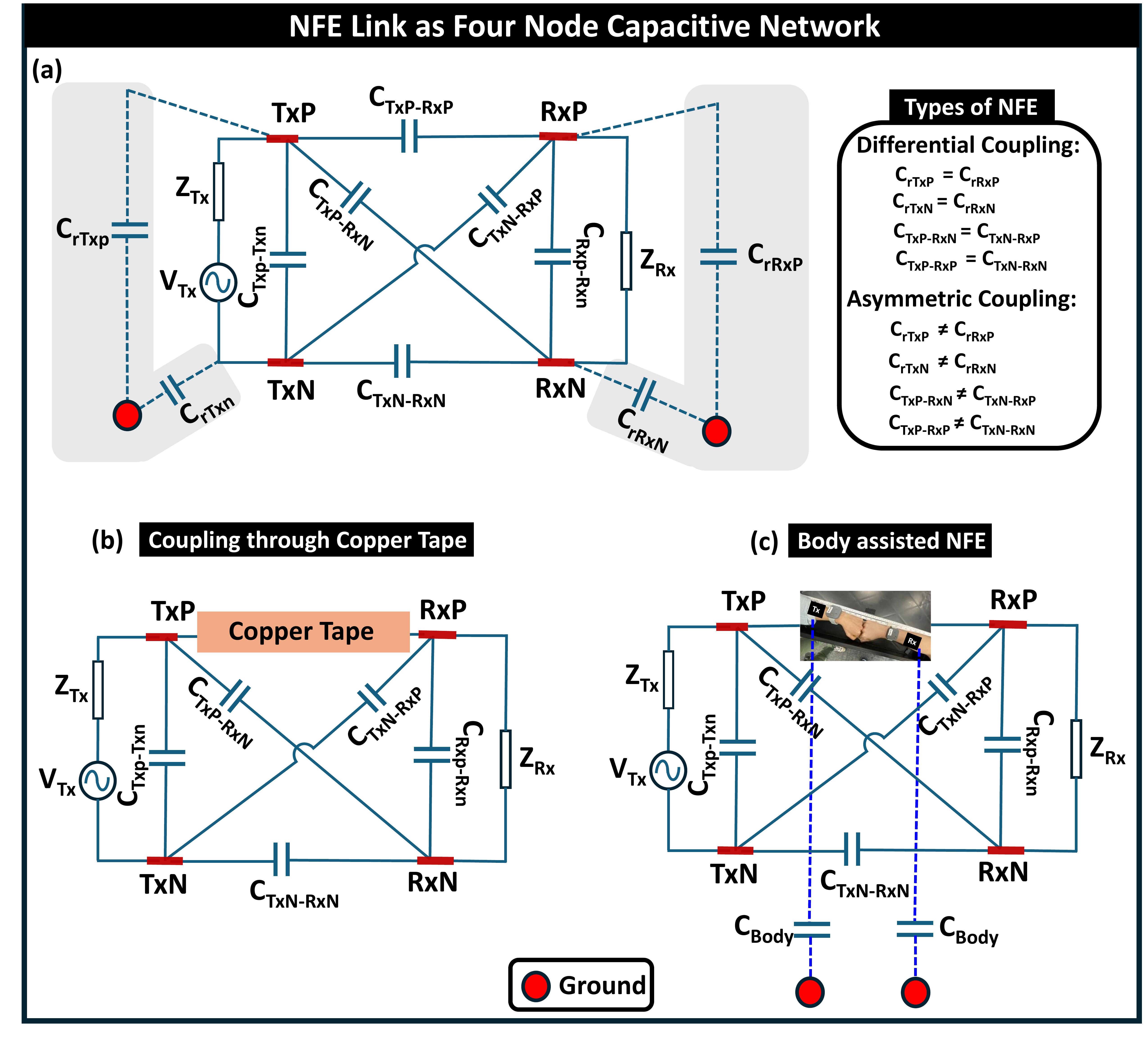}
\caption{NFE Link as Four Node Capacitive Network: (a) Different types of NFE, (b) Coupling through copper tape, (c) Body-assited NFE}
\label{fig:NFE_capacitive_network}
\end{figure}
\section{Near-Field Electric (NFE)}
Near Field Electric (NFE) communication—leverages strong, evanescent electric coupling—offers a fundamentally different trade-off: energy-efficient, spatially confined links that deliver high-speed data transfer without the broad radiation footprint of far-field systems \cite{das2019enabling}. This paper positions NFE as a practical bridge between ultra-short-range near-field techniques and conventional radio-frequency communications, unlocking new possibilities for dense, power-constrained networks. NFE’s appeal lies in three complementary strengths. First, its coupling mechanism concentrates energy in the immediate vicinity of the Tx and Rx, enabling transmit power orders of magnitude lower for a given data rate. Second, spatial confinement reduces mutual interference and enhances coexistence in crowded environments, making NFE attractive for Internet of Things (IoT), wearable, AR/VR, and industrial-sensing applications. Third, the physics of near-field electric interactions supports novel coupler and transceiver architectures that can trade complexity for lower consumed power, opening the way for compact, battery-friendly designs. Realizing these benefits requires overcoming distinct technical challenges: robust modulation and coding under strong spatial gradients. These practical transducer designs balance coupling strength with form-factor constraints, and safety and regulatory considerations unique to decimeter-range electric-field couplings.

Through experimental evidence and system-level evaluation, this work demonstrates that NFE at decimeter range can open up immense possibilities for energy-efficient next-generation, high-speed short-range communication.

It introduces Wi-R NFE E-field-based Transceiver (XA-NFE2001), which uses NFE communication to establish a contained electric-field link between devices, creating a private, hard-to-intercept data channel. By making physical proximity the primary security mechanism, it turns a simple touch or close placement into an intuitive authentication step. When devices are tapped together or brought into close range, they can automatically exchange data at a physically secure 5 Mbps, removing the need for complex software pairing. This contact-range method is far more energy-efficient than broadcasting radio signals, since it confines transmission to the immediate vicinity of the devices.

\section{Different modes of NFE}
This section presents different operational modes of NFE, with differential coupling and asymmetric coupling being two important approaches. Each has unique characteristics in terms of field distribution, received signal strength, and interference management.

Under quasi‑static, lumped‑element assumptions, NFE links can be conceptually modeled as a four-node capacitive network. These nodes are as follows: Tx signal ($\mathrm{TxP}$), Tx ground ($\mathrm{TxN}$), Rx signal ($\mathrm{RxP}$), and Rx ground ($\mathrm{RxN}$), shown in Fig. \ref{fig:NFE_capacitive_network}. Now, since the capacitive network is reciprocal, the mutual capacitances between the pair of nodes satisfy $C_{i-j}=C_{j-i}$ for every pair of nodes (i, j) and $C_{ri}$ denotes the self‑capacitance of node i to a global reference (true ground). With source $V_{Tx}$ being the applied input excitation between $\mathrm{TxP}$ and $\mathrm{TxN}$, the Rx measures $V_{Rx}$ between $\mathrm{RxP}$ and $\mathrm{RxN}$. The effective differential coupling ($\mathrm{C_{eff}}$) between the signal-ground electrodes of the Tx and Rx can be obtained by combining direct and cross terms:

\begin{equation}
    \centering
    \mathrm{C_{eff} =C_{TxP-RxP} + C_{\mathrm{TxN-RxN}}-C_{TxP-RxN} -C_{TxN-RxP}} 
\end{equation}

The coupling can be modeled as a series-capacitive path between the differential ports. The series impedance of coupling is $\mathrm{Z_C = \frac{1}{j \omega C_{eff}}}$. 
The differential capacitance seen across the Rx port is given by:
\begin{equation}
    \mathrm{C_{RR} =
C_{rRxP}
+
C_{rRxN} -2C_{RxP-RxN}}
\end{equation}

The Rx port sees the parallel combination of its input impedance and its differential capacitance, i.e.,
\begin{subequations}\label{main1:a}
\begin{equation}
\mathrm{Y_{L}(s)=s C_{RR}  + \frac{1}{Z_{Rx}}}
\label{eq5:main:a}
\end{equation}
\begin{equation}
\mathrm{Z_{L}(s)=\frac{1}{Y_{L}(s)} = \frac{1}{s C_{RR} + \dfrac{1}{Z_{Rx}}}}
\label{eq5:main:b}
\end{equation}
\end{subequations}

The simplified two-port loop yields a voltage divider between $\mathrm{Z_{Tx} + Z_C}$ and $\mathrm{Z_{L}}$ (assuming $\mathrm{Z_{Rx}}$ referenced to Rx differential port). The received voltage ($\mathrm{V_{Rx}}$) can be represented as 
\begin{equation}
    \centering
    \mathrm{V_{Rx} = \frac{Z_{L}}{(Z_{Tx} + Z_C + Z_{L})} \cdot V_{Tx}}
\end{equation}

Substituting $\mathrm{Z_C}$ and $\mathrm{Z_L}$ in the above expression of $\mathrm{V_{Rx}}$, we get 

\begin{equation}
    \centering
    \mathrm{H(j\omega) = \frac{V_{Rx}}{V_{Tx}}= \frac{1}{1 + Z_{Tx}\!\left(sC_{RR} + \frac{1}{Z_{Rx}}\right)
+ \frac{sC_{RR} + \frac{1}{Z_{Rx}}}{sC_{eff}}}}
\end{equation}
The low frequency limit of the transfer function can be set as $\mathrm{H\propto s C_{eff} Z_{Rx}}$. 

Assume Rx input sees thermal noise with one-sided power spectral density (PSD) $\mathrm{N_0}$ and bandwidth $\mathrm{B}$. Received signal power can be expressed as:

\begin{subequations}\label{main:a}
\begin{equation}
\mathrm{P_{Rx} = \lvert H(j\omega_0)\rvert^2 P_{Tx}}
\label{eq8:main:a}
\end{equation}
\begin{equation}
\mathrm{SNR= \frac{P_{Rx}}{N_0B}= \frac{\lvert H(j\omega_0)\rvert^2 P_{Tx}}{N_0 B}}
\label{eq8:main:b}
\end{equation}
\end{subequations}

\subsection{Differential Coupling with Symmetric Electrodes}
In differential coupling, communication relies on a pair of symmetric or similarly sized electrodes or conductors to transmit signals.  This balanced configuration creates an electric field that is confined between the electrodes, reducing radiation into the far field. Because the fields are contained mainly, differential coupling offers better immunity to external noise and interference. The system's symmetry also minimizes common-mode signals, helping achieve stable, reliable data transfer. The operation of differential coupling can be explained from the mathematical formulation presented above: during symmetry, so cross terms cancel in $\mathrm{C_{eff}}$. Hence, this requires identical electrodes and balancing networks.

This modality is often used in secure or short-range communication systems where maintaining signal integrity is critical. Additionally, differential coupling tends to be more energy-efficient because the field is concentrated in the near-field, reducing unnecessary losses. The direct coupling can be further categorized as: (a) Direct Coupling (Air), (b) Coupling through dielectric (doors, glass, etc.). The scenario of direct coupling occurs when devices capacitively couple through air due to mutual capacitance between their electrodes. On the contrary, coupling through a dielectric occurs when the relative permittivity of the intervening medium enhances the coupled signal strength.   
\subsection{Asymmetric for Extended Range}
In contrast, asymmetric coupling uses an unbalanced configuration in which one electrode is larger than the other. This creates an electric field distribution that extends more widely into the surrounding medium than differential coupling, thereby achieving greater communication coverage \cite{sarkar2025human}. 
While this approach can achieve longer communication distances in the near-field region, it is more susceptible to external interference because of its less confined field distribution. Similar to symmetric coupling, the scenario of asymmetric coupling can be formulated as: one of the cross-coupling capacitances (e.g., $\mathrm{C_{TxP-RxN}}$ or $\mathrm{C_{TxN-RxP}}$) dominates. However, this causes an increase in $\mathrm{C_{eff}}$, leading to larger common-mode terms and increased susceptibility to external noise and potential far-field leakage.

Asymmetric coupling can be advantageous in low-cost or compact systems where design simplicity outweighs the need for maximum noise immunity. However, designers must carefully manage the resulting common-mode signals and potential leakage into the far field to avoid performance degradation. The scenarios of asymmetric coupling can be classified under the following categories: 

(a) \textbf{Signal Plate extended} e.g., Coupling through conductors, shown in Fig. \ref{fig:NFE_capacitive_network} (b), (b) \textbf{Ground plate extended}: e.g., Body-assisted NFE with extended coverage, shown in Fig. \ref{fig:NFE_capacitive_network} (c). 

\begin{figure}[ht]
\centering
\includegraphics[width=0.48\textwidth]{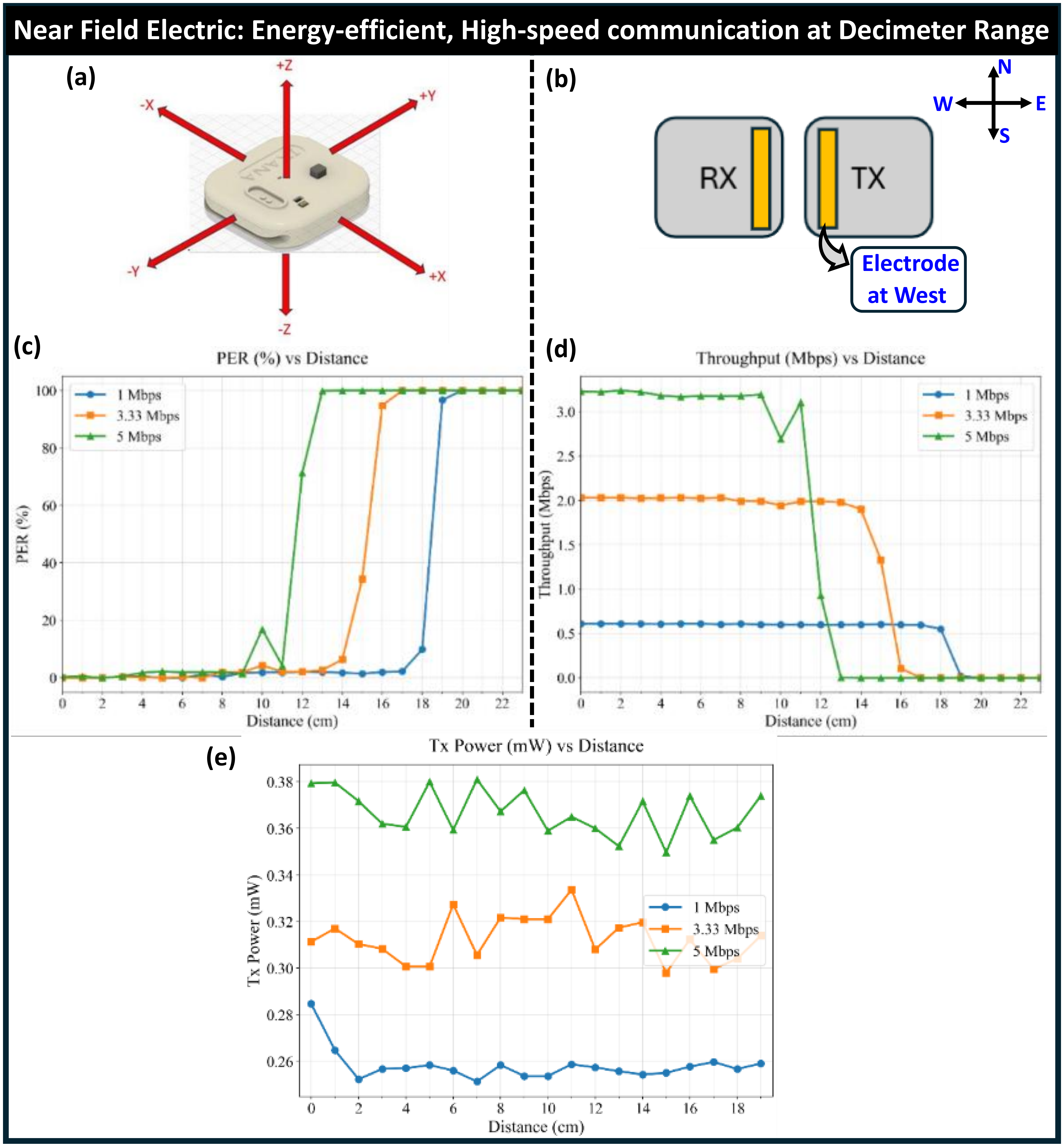}
\caption{Near Field Electric (NFE) Communication enabling energy-efficient, high-speed communication at decimeter Range: (a) NFE devices feature electrodes positioned along the -Z axis, facing downward. (b) Bottom View of the Devices: This view illustrates the electrode positions and the optimal communication configuration, highlighting the ideal settings for Packet Error Rate (PER) and throughput results. The optimal throughput is over 600 kbps at 1 Mbps, over 2 Mbps at 3.33 Mbps, and over 3 Mbps at 5 Mbps. (c) The variation of Packet Error Rate (PER) with transmission distance is presented. (d) The variation of throughput with transmission distance is shown. (e) An optimal communication configuration with the best settings for transmission (Tx) and reception (Rx) power results is provided.}
\label{fig:Optimized Performance Results}
\end{figure}

\section{Measurement Setup}
This section presents the details of the communication devices used for NFE-based measurement setup. The devices are made with Wi-R NFE E-field Transceiver (XA-NFE2001) chip from Ixana. Its operating voltage lies in the range of 1.7 V–2.7 V (1.8 V/2.5 V compliant). It has the ability to support an adjustable carrier/data rate up to 5 Mbps. It has an impressively lower power consumption ($<$1 mW per transceiver at 5 Mbps). Moreover, the Software Development Kit (SDK) of the transceiver has feature to support simultaneous communication with up to 16 devices. The two NFE devices are of size 48 mm x 44 mm x 13 mm. 

\section{Measurement Procedure $\&$ Results}
The experiments involved a pair of NFE devices configured as a Tx and a Rx, as shown in Fig. \ref{fig:Optimized Performance Results} (a). The NFE devices feature a rectangular electrode aligned along the (-Z axis), and the experiments were conducted with the electrode facing downward, illustrated in Fig. \ref{fig:Optimized Performance Results} (b).

This section provides a detailed evaluation of the communication performance of the NFE devices under various testing configurations, including changes in orientation, electrode geometry, long-distance transmission, and on-body operation. Performance was measured at 1 Mbps, 3.33 Mbps, and 5 Mbps across different distances and during on-body trials. The tests focused on key metrics, such as Packet Error Rate (PER) and throughput, to assess the reliability of the wireless link across a range of experimental conditions. Multiple trials were conducted to ensure consistent results and identify performance trends across different test scenarios.

The results for the optimal communication configuration are presented in Fig. \ref{fig:Optimized Performance Results}(c, d, e). The variation in packet error rate with transmission distance for the optimal configuration is captured in Fig. \ref{fig:Optimized Performance Results}(c). The optimal throughput achieved is $>$600 kbps at 1 Mbps, $>$2 Mbps at 3.33 Mbps, and $>$3 Mbps at 5 Mbps, as shown in Fig. \ref{fig:Optimized Performance Results} (d). Additionally, Fig. \ref{fig:Optimized Performance Results}(e) illustrates the variation in power consumption at the Tx for the three selected data rates.

\begin{figure}[ht]
\centering
\includegraphics[width=0.48\textwidth]{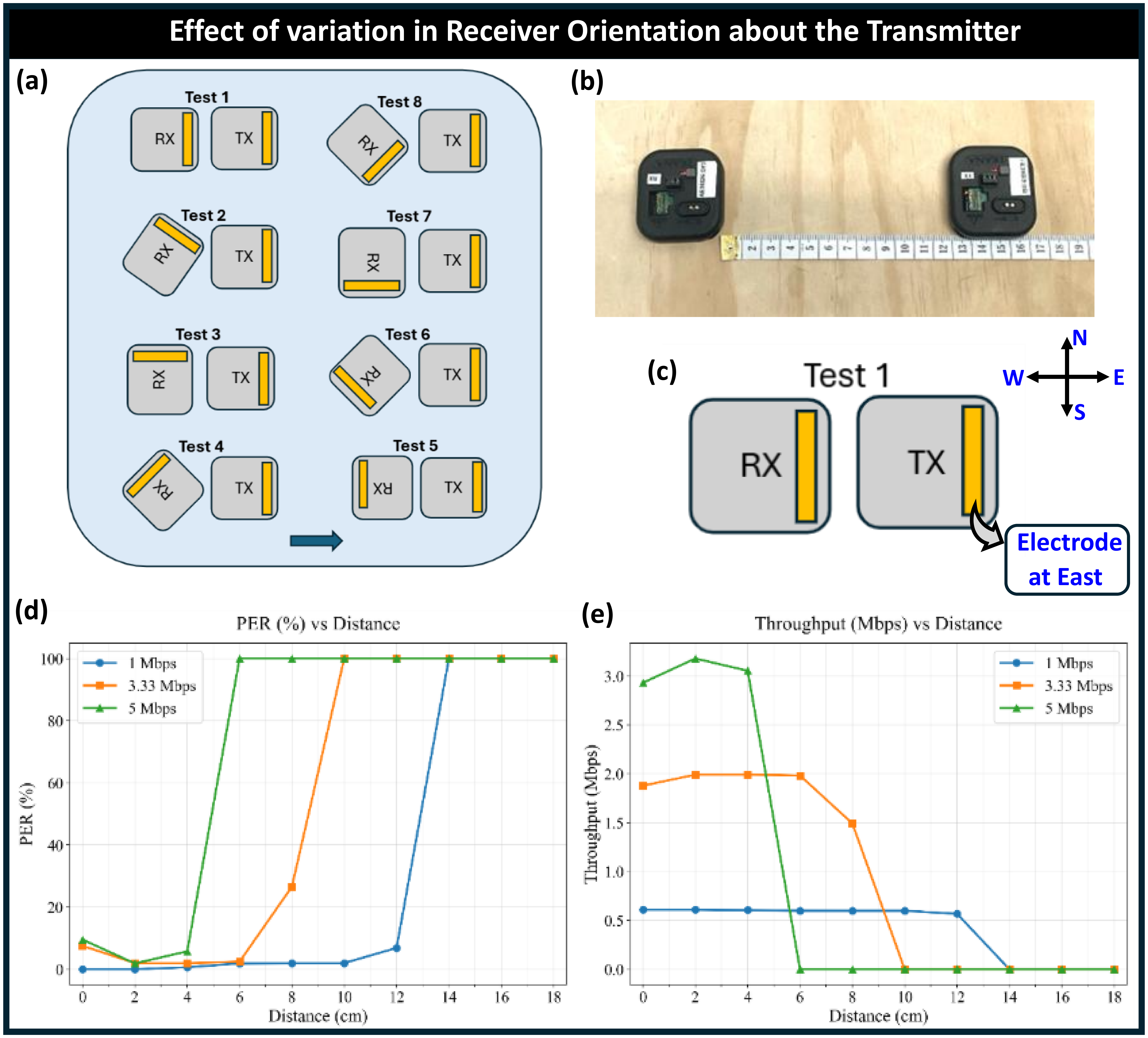}
\caption{Effect of Variation in Rx Orientation with the Tx Electrode Positioned to the Right (East): (a) Various orientations for Testing Configuration 1, where the Rx has undergone 45$^\circ$ rotations.
(b) Overview of the test setup for Configuration 1 of Test 1.
(c) Bottom view of the devices, highlighting the electrode position for Configuration 1 of Test 1.
(d) Variation in packet error rate with distance.
(e) Variation in throughput with distance.}
\label{fig:Electrode_Orientation_Part1}
\end{figure}

\begin{figure}[ht]
\centering
\includegraphics[width=0.48\textwidth]{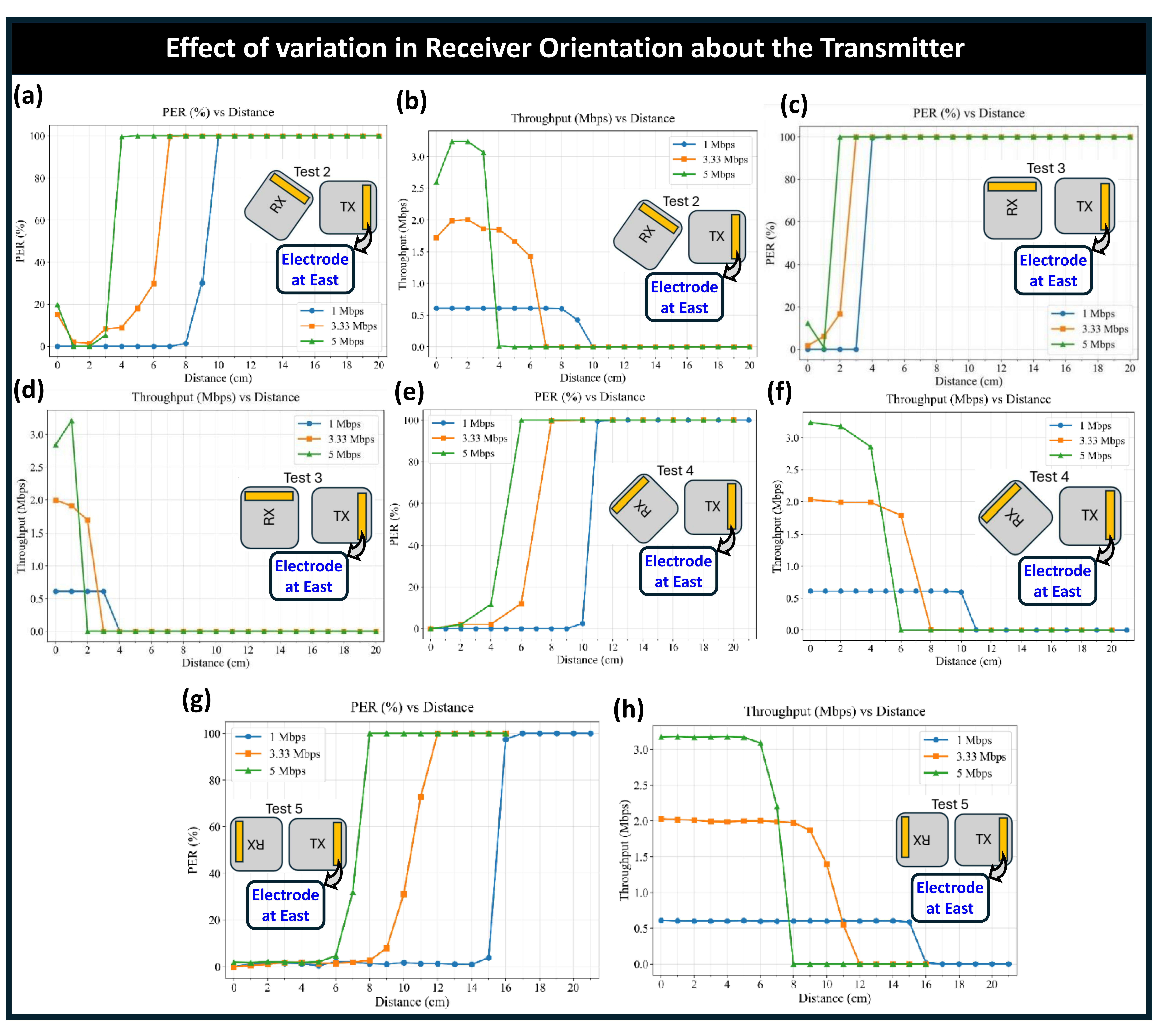}
\caption{Configuration 1 Test 2:(a) PER variation, (b) Throughput variation with distance; Configuration 1 Test 3:(c) PER variation, (d) Throughput variation with distance; Configuration 1 Test 4:(e) PER variation, (f) Throughput variation with distance; Configuration 1 Test 5:(g) PER variation, (h) Throughput variation with distance.
Inset figure in each case illustrates the electrode orientation.}
\label{fig:Electrode_Orientation_Part2}
\end{figure}

\begin{figure}[ht]
\centering
\includegraphics[width=0.48\textwidth]{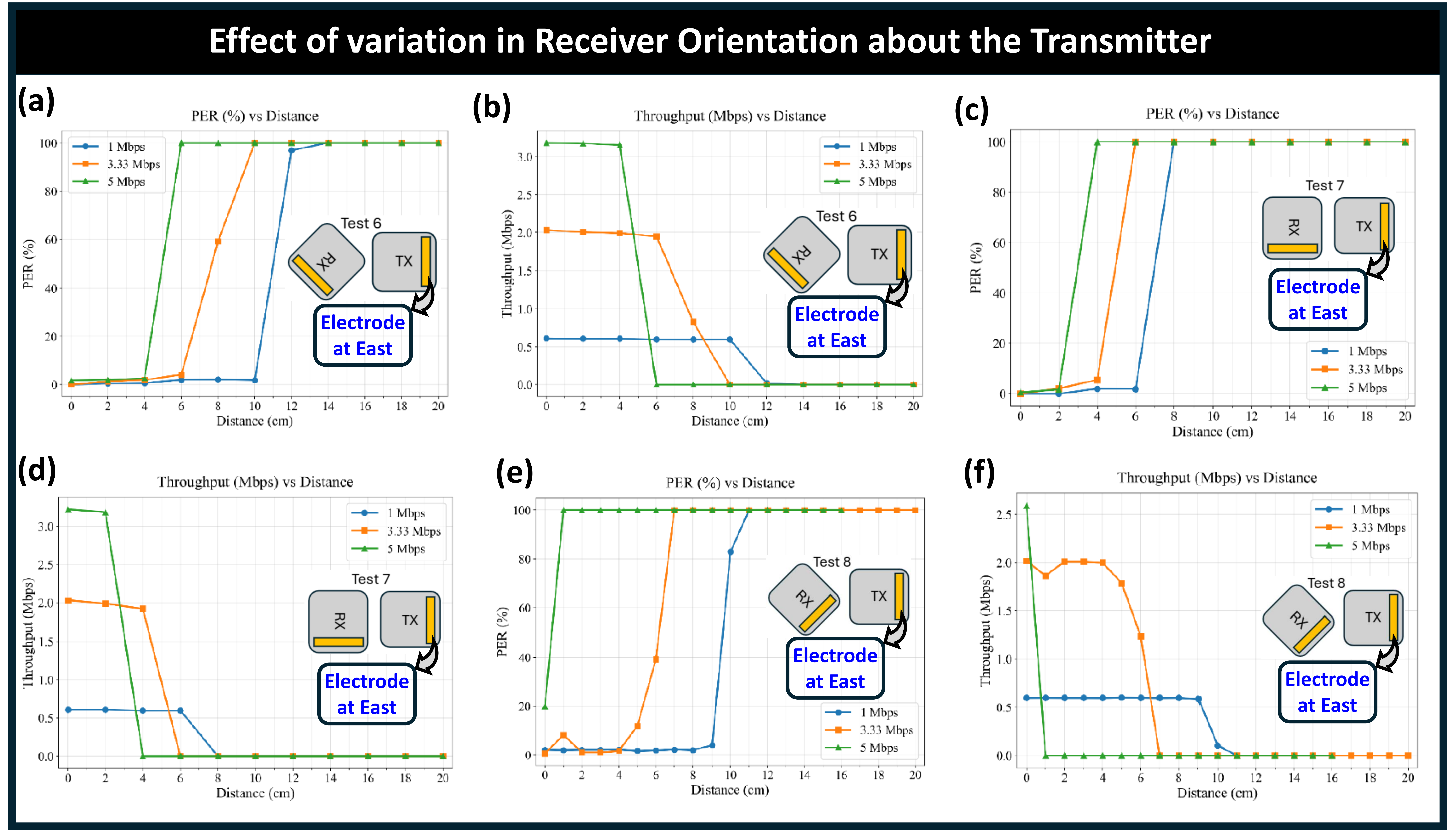}
\caption{Configuration 1 Test 6:(a) PER variation, (b) Throughput variation with distance; Configuration 1 Test 7:(c) PER variation, (d) Throughput variation with distance; Configuration 1 Test 8:(e) PER variation, (f) Throughput variation with distance.
Inset figure in each case depict the electrode orientation.}
\label{fig:Electrode_Orientation_Part3}
\end{figure}

\begin{figure}[ht]
\centering
\includegraphics[width=0.48\textwidth]{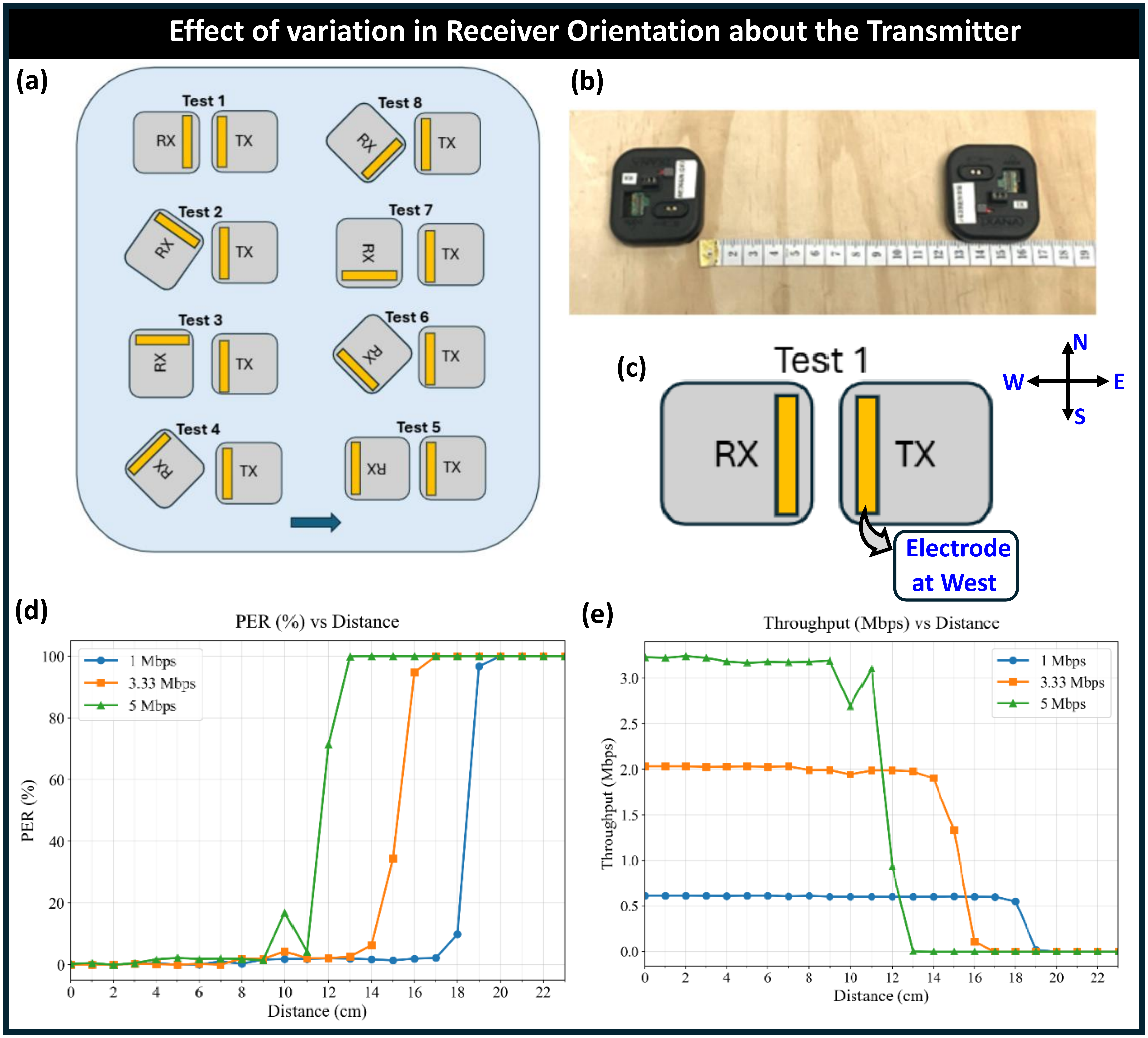}
\caption{Effects of Variation in Rx Orientation with the Tx Electrode Positioned on the Left (West):
(a) Different orientations for testing Configuration 2, where the Rx is rotated at 45$^\circ$.  
(b) Reference for the test setup of Configuration 2 in Test 1.  
(c) Bottom view of the devices: electrode positioning for Configuration 1 of Test 1.  
(d) Variation in packet error rate.  
(e) Variation in throughput with distance.}
\label{fig:Electrode_Orientation_Part4}
\end{figure}

\begin{figure}[ht]
\centering
\includegraphics[width=0.48\textwidth]{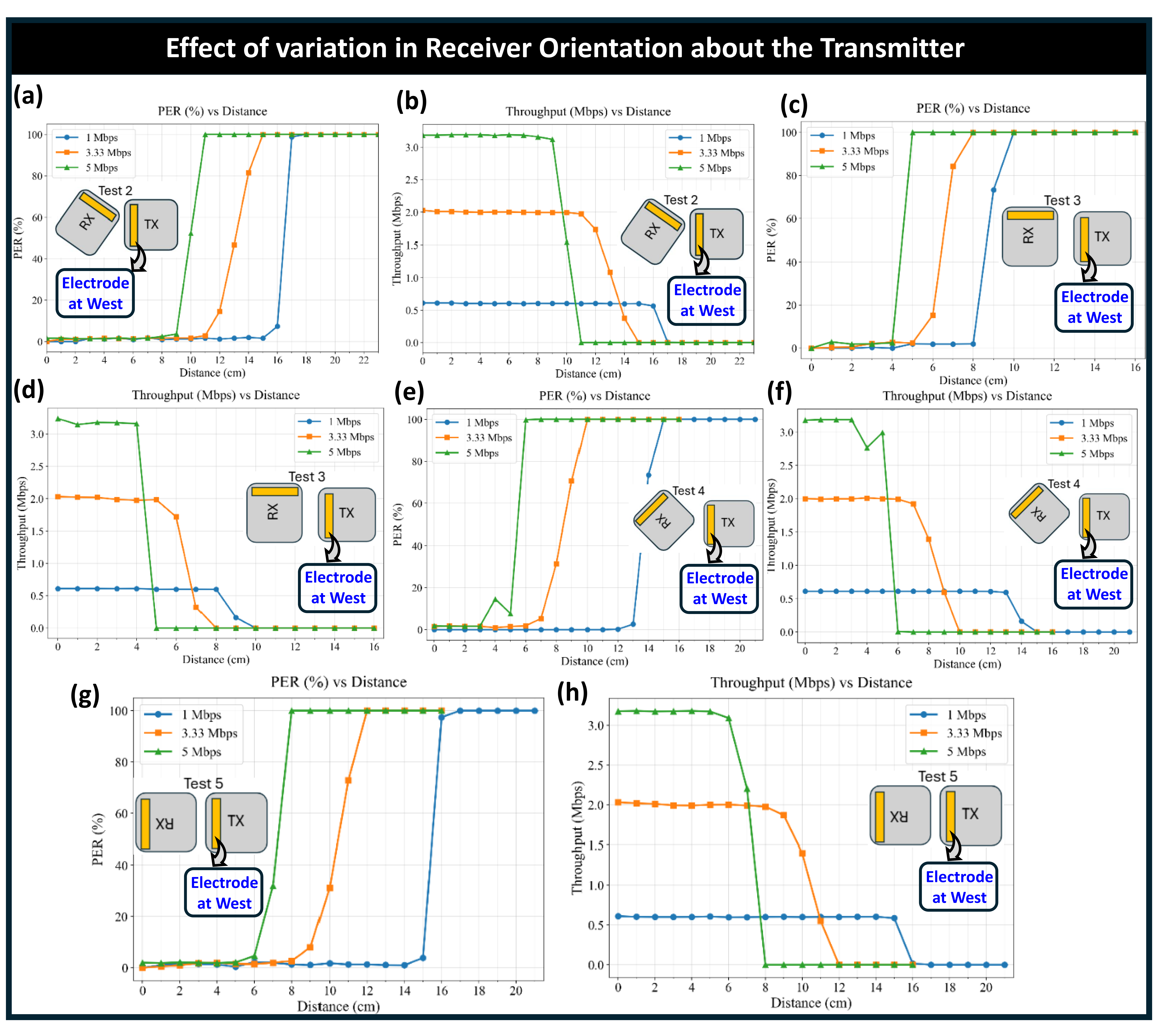}
\caption{Configuration 2 Test 2:(a) PER variation, (b) Throughput variation with distance; Configuration 2 Test 3:(c) PER variation, (d) Throughput variation with distance; Configuration 2 Test 4:(e) PER variation, (f) Throughput variation with distance; Configuration 2 Test 5:(g) PER variation, (h) Throughput variation with distance.
Inset figure in each case illustrates the electrode orientation.}
\label{fig:Electrode_Orientation_Part5}
\end{figure}

\begin{figure}[ht]
\centering
\includegraphics[width=0.48\textwidth]{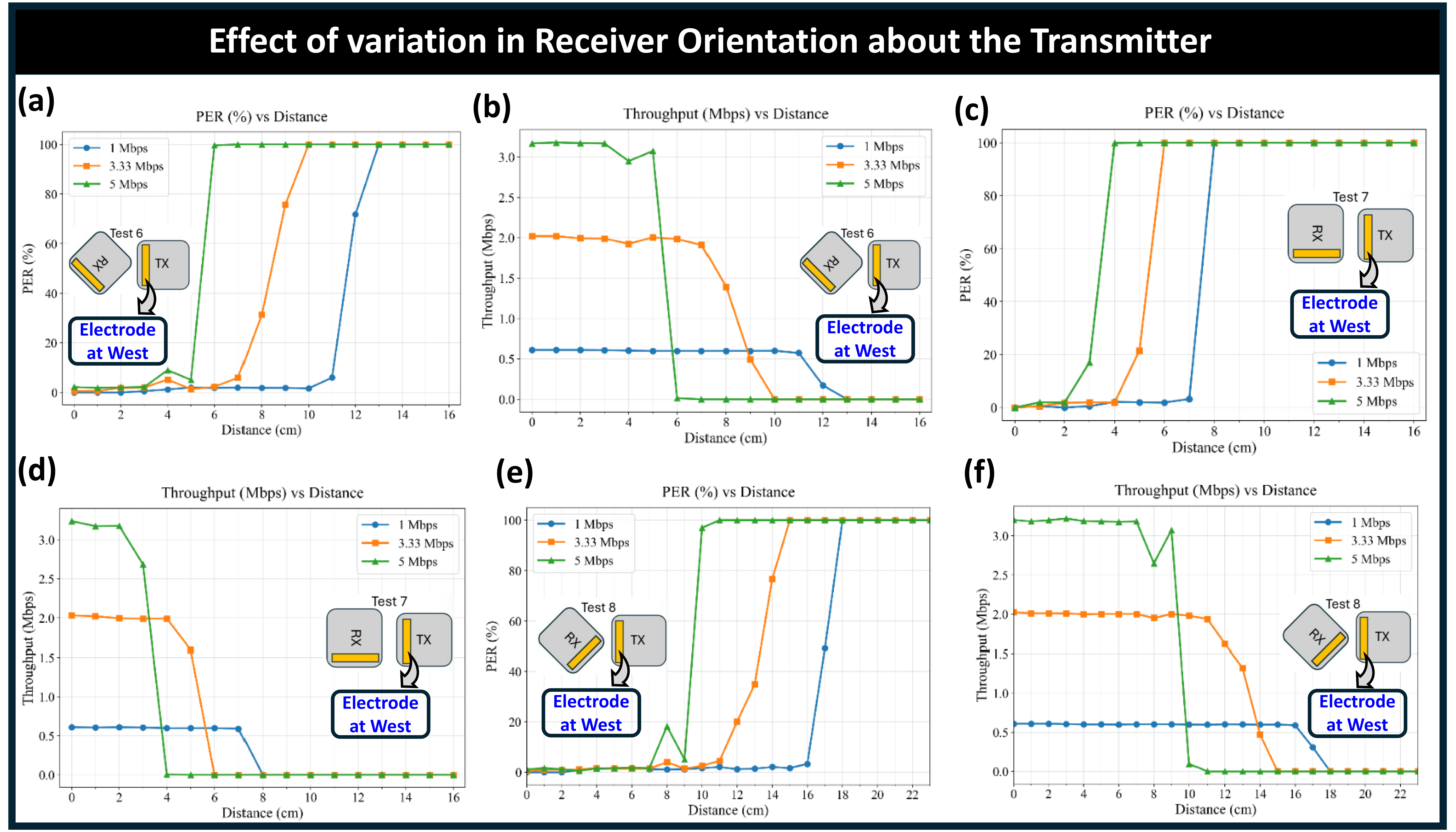}
\caption{Configuration 2 Test 6:(a) PER variation, (b) Throughput variation with distance; Configuration 2 Test 7:(c) PER variation, (d) Throughput variation with distance; Configuration 2 Test 8:(e) PER variation, (f) Throughput variation with distance.
Inset figure in each case depict the electrode orientation.}
\label{fig:Electrode_Orientation_Part6}
\end{figure}

\begin{figure}[ht]
\centering
\includegraphics[width=0.48\textwidth]{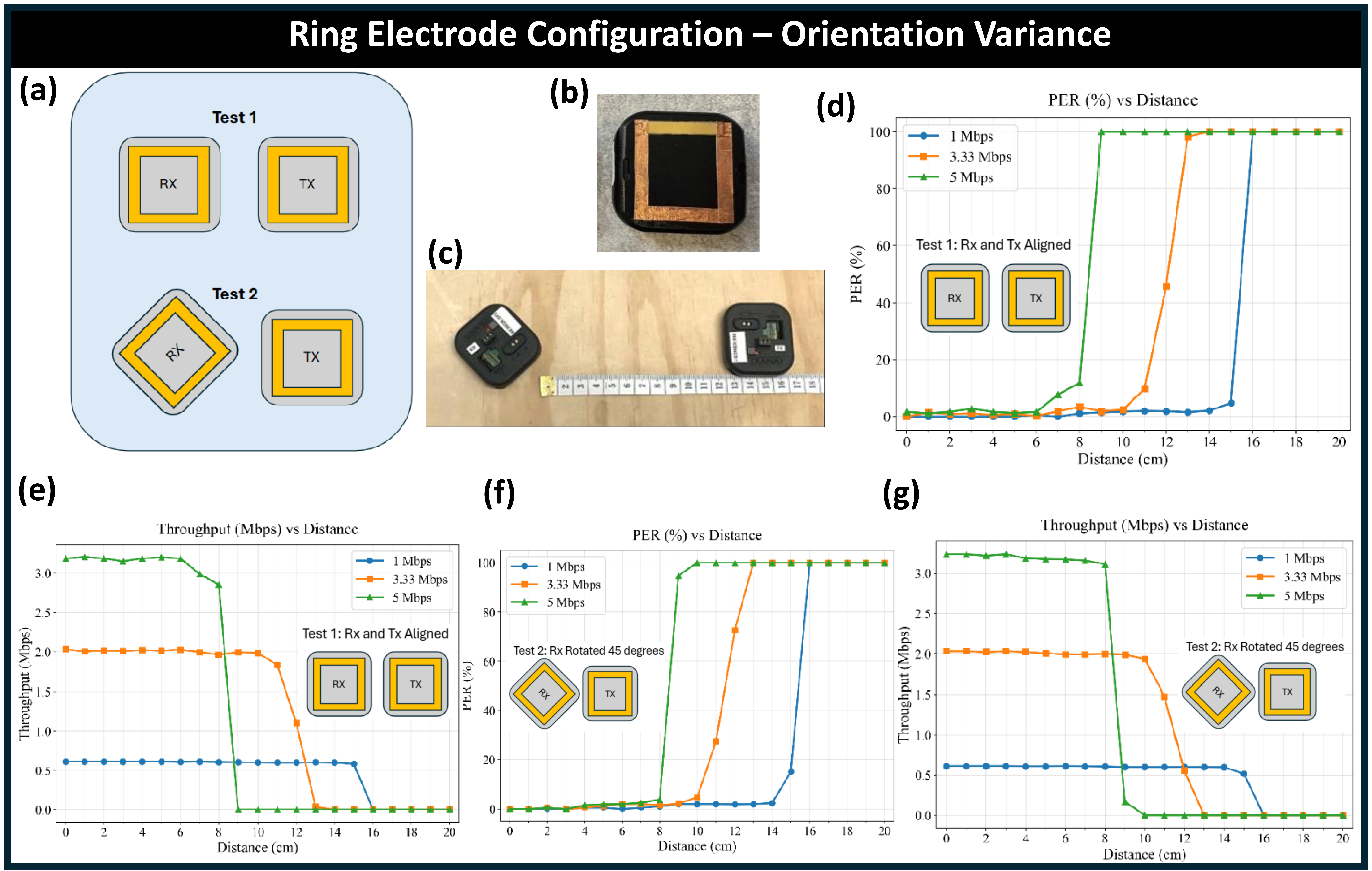}
\caption{Analysis of Orientation Variance with Ring Electrode Configuration: (a) Configuration 3 features symmetrical electrodes and includes two testing orientations. 
(b) An image of the symmetrical electrode is shown on the underside of the NFE device. 
(c) Reference for the test setup of Configuration 3, Test 1: 
(d) Variation in PER (Packet Error Rate).
(e) Variation in throughput.
For Test 2: 
(f) Variation in PER.
(g) Variation in throughput in relation to the transmission distance.}
\label{fig:Electrode_Orientation_Ring}
\end{figure}

\subsection{Electrode's Orientation Variance with Tx Electrode on Right}
During testing, both devices are positioned flat on a non-conductive surface with their electrodes facing downward and touching the surface. The Tx device remains fixed, with its electrode oriented toward the east, while the Rx device is rotated counterclockwise by 45$^\circ$ for each successive test. 
A total of eight orientation tests is performed, completing a full 360$^\circ$ rotation of the Rx device.  
The signal electrode of both the NFE devices (Tx and Rx) is facing downwards, positioned along the Z axis. Fig. \ref{fig:Electrode_Orientation_Part1}(a) illustrates the eight testing orientations for configuration 1. The reference of the test setup of configuration 1, Test 1, is presented in Fig. \ref{fig:Electrode_Orientation_Part1}(b).
For each data rate, the Tx device was moved in 1 cm increments away from the fixed receiving (Rx) device. At each distance, three consecutive runs were conducted, and the average results over the three runs are plotted. This process was repeated for all three data rates (1 Mbps, 3.33 Mbps, and 5 Mbps) and for each of the eight Rx orientations (Tests 1–8). During all tests, the devices were configured for continuous transmission and reception, and measurements were conducted in a controlled, no-interference environment. The variation in PER and throughput with transmission distance for test 1 is captured in Fig. \ref{fig:Electrode_Orientation_Part1} (d) and (e), respectively. Test 1 showed transmission ranges of 13 cm, 9 cm, and 5 cm at 1 Mbps, 3.33 Mbps, and 5 Mbps, respectively. 

The performance of the NFE devices in terms of their variation in PER and throughput under the remaining test conditions is summarized via the results from Test 2-5, illustrated in Fig. \ref{fig:Electrode_Orientation_Part2} $\&$ from Test 6-8 shown in Fig. \ref{fig:Electrode_Orientation_Part3}. 

Results show that communication performance between transmitting and receiving NFE devices depends on both data rate and Rx orientation. As expected, the effective communication range decreases as the data rate increases. The optimum throughput achieved at a 5 Mbps data rate is $>$3 Mbps, achievable for all orientations and shorter channel lengths. As channel length increases, the throughput decreases. Across all orientations, the 1 Mbps configuration consistently achieved the higher communication distance, maintaining acceptable PER levels (below 30$\%$) up to approximately 10–14 cm in most cases. The 3.33 Mbps configuration showed moderate performance, with the 30$\%$ PER threshold typically reaching between 6–10 cm, depending on orientation. At the highest data rate of 5 Mbps, reliable communication was generally limited to 1–6 cm, reflecting the expected trade-off between throughput and range in near-field communication systems. Specific orientations, such as Test 1, produced consistently higher ranges across all data rates, indicating stronger communication between the Tx and Rx electrodes. Test 1 is also when the electrodes are closest together. These variations are consistent with near-field device behavior, where electrode orientation and positioning directly affect transmission strength. 

\subsection{Electrode's Orientation Variance with Tx Electrode on Left}
Here, the Tx device remains fixed with its electrode facing west, while the Rx device is rotated counterclockwise by 45° for each successive test. A total of eight orientation tests are performed, completing a full 360$^\circ$ rotation of the Rx device, as in the previous scenario. For each orientation, continuous transmission and reception are maintained while measuring communication performance across the distance range at each data rate. 

A total of 8 tests with Rx device rotating 45 degrees counterclockwise and Tx electrode remaining fixed on the west side, shown in Fig. \ref{fig:Electrode_Orientation_Part4} (a). The reference for the testing setup of configuration 2, Test 1, is presented in Fig. \ref{fig:Electrode_Orientation_Part4} (b). The electrodes of both NFE devices (Rx and Tx) are facing downward, aligned along the –Z axis, illustrated in Fig. \ref{fig:Electrode_Orientation_Part4} (c). The variation in PER and throughput for the selected data rates is captured in Fig. \ref{fig:Electrode_Orientation_Part4} (d) and Fig. \ref{fig:Electrode_Orientation_Part4} (e), respectively.

The PER and throughput performance under the remaining test conditions are summarized via the results from Test 2-5, presented in Fig. \ref{fig:Electrode_Orientation_Part5} $\&$ from Test 6-8, shown in Fig. \ref{fig:Electrode_Orientation_Part6}.

The results from Testing Configuration 2 (Tx electrode facing west) indicate strong and consistent communication performance between the NFE Tx and Rx devices, with both data rate and orientation continuing to play a significant role in link range. The optimum throughput achieved at a 5 Mbps data rate is $>$3 Mbps, achievable for all orientations and shorter channel lengths. As channel length increases, the throughput decreases. In contrast to Configuration 1 (Tx electrode facing east), the west-facing Tx orientation generally provided improved coupling and extended communication ranges across multiple tests. At 1 Mbps, the effective communication distance before exceeding 30$\%$ PER ranged between 1–19 cm, with several orientations, particularly Tests 4, 5, 6, and 8, demonstrating ranges above 12 cm. At 3.33 Mbps, the communication range was typically 6–15 cm, whereas at 5 Mbps it remained stable at 4–12 cm, showing better high-rate performance than Configuration 1. 

Overall configuration 2 demonstrates enhanced range and link reliability across all data rates, validating that electrode orientation plays a key role in optimizing near-field coupling and ensuring stable communication performance across the tested conditions. 

\subsection{Orientation Variance with Ring Electrode Configuration}
To understand how the electrode geometry and orientation affect transmission range and link stability at data rates of 1 Mbps, 3.33 Mbps, and 5 Mbps, the following test, presented in configuration 3, is conducted, with the Tx and receiving (Rx) NFE devices when equipped with symmetrical square electrodes.   

Both the Tx and Rx devices feature square-shaped electrodes, positioned at the bottom surface of each device, shown in Fig. \ref{fig:Electrode_Orientation_Ring}. The devices are placed flat on a non-conductive surface, with the electrodes facing downward and aligned along the –Z axis, as shown in Fig. \ref{fig:Electrode_Orientation_Ring} (b, c).

The devices were tested at the three selected data rates. For each data rate, the Tx device was moved in 1 cm increments away from the fixed Rx device. At each distance, three consecutive runs were performed to ensure consistency and repeatability. Tests 1 and 2 were each conducted across all three data rates under continuous transmission and reception, in a controlled, interference-free environment. The test scenarios are illustrated in Fig. \ref{fig:Electrode_Orientation_Ring} (a).In Test 1, the Tx and Rx electrodes are perfectly aligned side by side, maintaining complete surface symmetry between the square electrodes. In Test 2, the Rx electrode is rotated 45$^\circ$ counterclockwise relative to the Tx electrode to evaluate the effect of a rotated alignment on near-field communication performance.

The PER and throughput performance from tests 1 and 2 are presented in Fig. \ref{fig:Electrode_Orientation_Ring} (d, e) and Fig. \ref{fig:Electrode_Orientation_Ring} (f, g), respectively. Results show that symmetrical square electrodes provide consistent, stable communication performance across all data rates tested. The optimum throughput achieved at a 5 Mbps data rate is $>$3 Mbps, achievable for all orientations and shorter channel lengths. As channel length increases, the throughput decreases. In Test 1, where the Tx and Rx electrodes were perfectly aligned, the 30$\%$ PER threshold was approximately 16 cm at 1 Mbps, 12 cm at 3.33 Mbps, and 9 cm at 5 Mbps. In Test 2, where the Rx electrode was rotated 45$^\circ$ counterclockwise, the communication ranges remained nearly identical at 16 cm, 11 cm, and 9 cm at the respective data rates. 

Compared to previous configurations, the symmetrical square electrode design demonstrates greater orientation tolerance and improved stability in PER and throughput, making it a practical design for robust near-field communication. 

\subsection{Extended-Range using Conductor-coupling}
\begin{figure}[ht]
\centering
\includegraphics[width=0.48\textwidth]{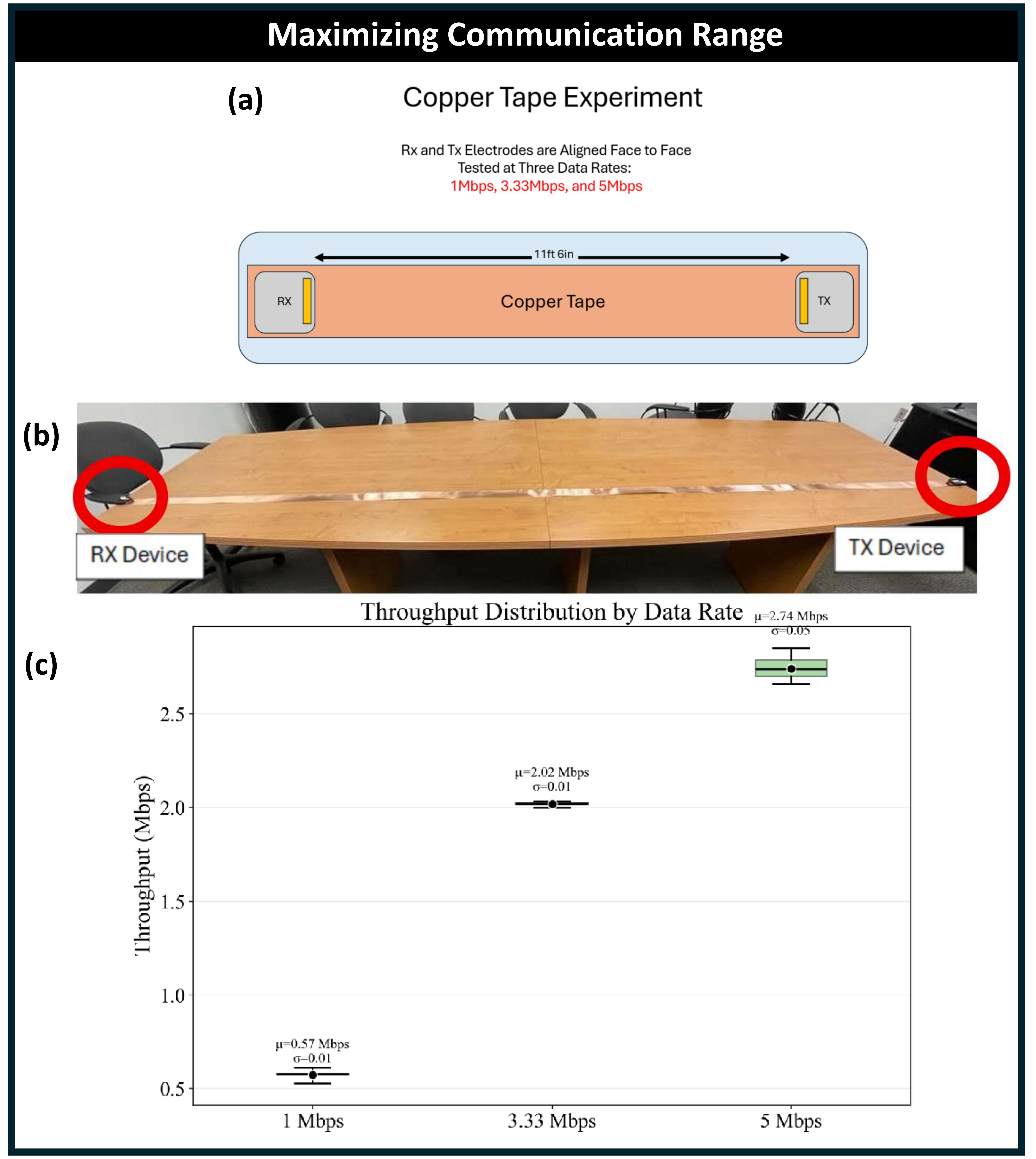}
\caption{Extended-range using Conductor-coupling: (a) Copper taper configuration tested at three specific data rates, (b) Actual test setup of a pair of NFE devices placed on continuous copper tape, and (c) Throughput distribution for various data rates.}
\label{fig:Conducting_Tape}
\end{figure}

To assess the communication performance of the NFE pair over an extended transmission distance, a conductive medium such as a commercially available, conductive double-sided copper tape is used to facilitate near-field coupling, presented in the test Configuration 4, shown in Fig. \ref{fig:Conducting_Tape} (a). Next, we determine how PER and throughput behave at long distances when the devices are separated well beyond the typical operating range, across the selected data rates of 1 Mbps, 3.33 Mbps, and 5 Mbps. 

A pair of NFE devices is placed on the continuous copper tape measuring approximately 11 ft 6 in ($\sim$3.5 meters) in length between them, illustrated in Fig. \ref{fig:Conducting_Tape} (b). The copper tape served as a conductive path to study long-distance communication effects. Both the Tx and Rx devices were positioned flat on the tape, with their electrodes facing downward and oriented face-to-face, similar to Configuration 2, Test 1. The devices were fixed at opposite ends of the tape to establish the maximum achievable communication separation. 

\begin{figure}[ht]
\centering
\includegraphics[width=0.48\textwidth]{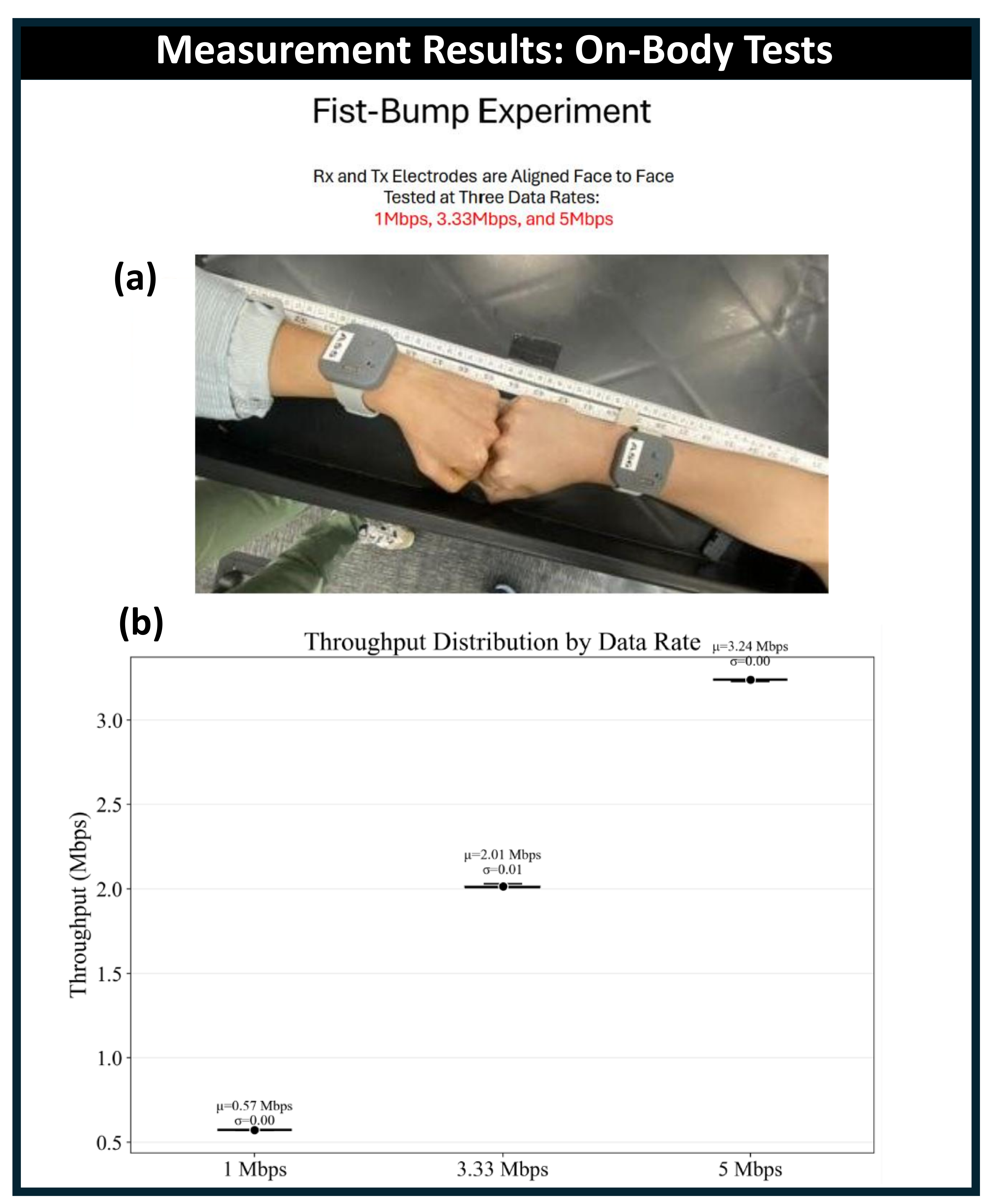}
\caption{On-Body Tests with regular NFE devices: (a) Fist Bump Experiment with a pair of NFE devices worn as watches and tested at three data rates, (b) Test setup of NFE pair as watches for fist bump experiment, fist to fist contact, (c) Throughput distribution for different data rates.}
\label{fig:On-Body_Tests}
\end{figure}

For each data rate, the devices were configured for continuous transmission and reception while PER and throughput were recorded. A total of 50 runs were performed for each test. Results of throughput distribution for the three selected data rates are shown in Fig. \ref{fig:Conducting_Tape} (c). Unlike previous configurations, the separation between the Tx and Rx was constant at 11 ft 6 in ($\sim$350 cm), allowing evaluation of signal propagation over an extended conductive medium. The tests were performed in a controlled, low-interference environment to ensure data accuracy. Results show that the data rate increased from 1 Mbps to 5 Mbps, and the normal Packet Error Rate (PER) remained below 30$\%$ (5.8$\%$ at 1 Mbps, 12.5$\%$ at 3.33 Mbps, and 29$\%$ at 5 Mbps). Correspondingly, throughput increased with data rate, reaching 0.55 Mbps, 1.6 Mbps, and 2 Mbps, respectively, for data rates of 1 Mbps, 3.33 Mbps, and 5 Mbps. Overall, the conductive tape successfully enabled communication beyond the typical operating range. 

\begin{figure}[ht]
\centering
\includegraphics[width=0.48\textwidth]{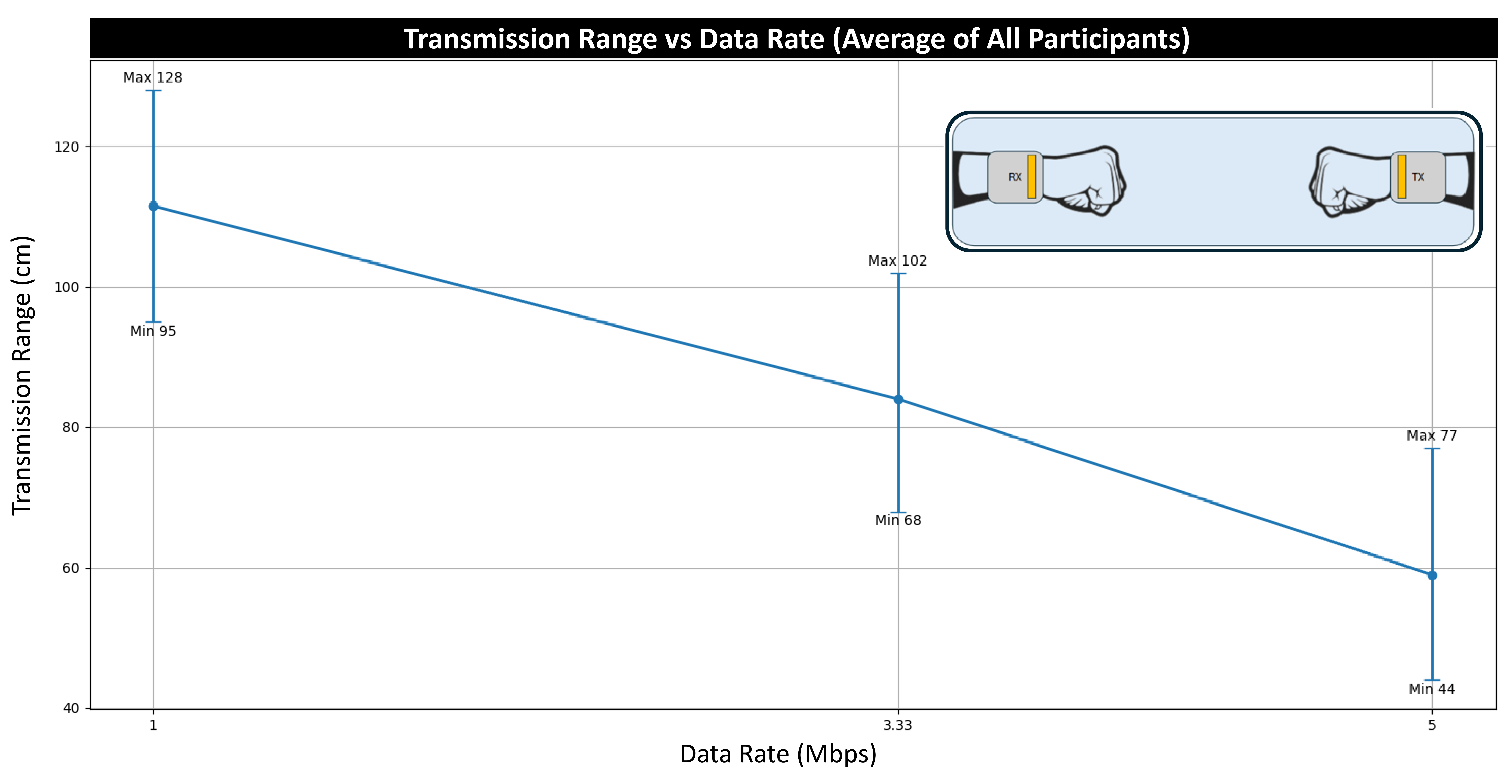}
\caption{Extended Communication Coverage using Range-enhanced NFE devices along with Body-coupling: A pair of users wore watches, and the communication range was identified between three pairs of users. Schematic diagram (inset) shows a pair of users with Range-enhanced NFE devices during the Fist-bump scenario.}
\label{fig:On-Body_Results}
\end{figure}

\subsection{Extended-range using Body-Coupling}
The transmission range of NFE can be further extended by utilizing the electrical conductivity of the human body \cite{maity2018bio, das2019enabling, datta2021advanced, yang2022physically}, as demonstrated in the previous scenario. Fig. \ref{fig:On-Body_Tests} illustrates the performance of the NFE-based wireless link using single electrodes at both the Tx and Rx in test configuration 5. Measurements were taken indoors with two seated users to ensure consistent device placement; wrist-worn NFE devices acted as the Tx and Rx, with the electrode orientation corresponding to configuration 2 from test 1. During the initial contact, represented by the fist-bump in Fig. \ref{fig:On-Body_Tests}(a), throughput was recorded at three different data rates: 1 Mbps, 3.33 Mbps, and 5 Mbps. Multiple runs were conducted to ensure consistency and reproducibility of the results. The average throughputs for the short-channel scenarios were found to be 0.57 Mbps at 1 Mbps, 2.01 Mbps at 3.33 Mbps, and 3.24 Mbps at 5 Mbps, as shown in Fig. \ref{fig:On-Body_Tests} (b). 

Using device design modifications and further range-enhancement techniques, in specific scenarios, a meter-scale communication may be established, as shown in Fig. \ref{fig:On-Body_Results}. It is observed that using Range-
enhanced NFE devices along with Body-coupling, communication coverage up to the meter scale at a selected data rate of 1 Mbps. At an increased data rate to 5 Mbps, the link can still support $>$0.5 m of wireless coverage. Experiments were repeated for 3 pairs of humans, and results are shown through error bars. Hence, depending upon the application requirements, the users can leverage the reconfigurable transmission range feature of proposed technique.

\begin{figure}[ht]
\centering
\includegraphics[width=0.48\textwidth]{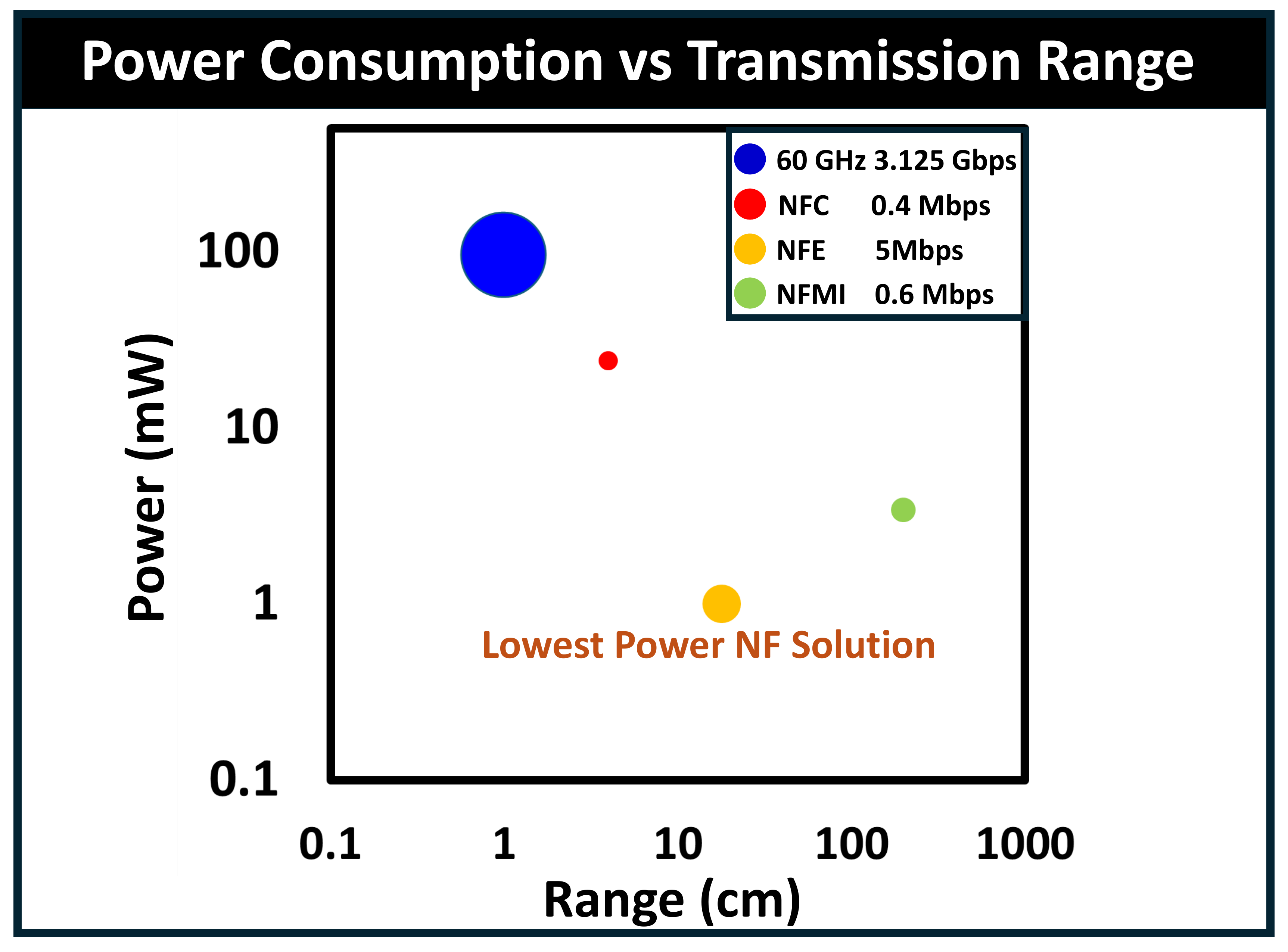}
\caption{Comparison of Performance Metrics: Power vs Transmission Range for different near field techniques like NFC, NFMI, mm-wave and the proposed NFE. Size of the marker representing the supported data rate.}
\label{fig:Performance Comparison}
\end{figure}

\begin{figure}[ht]
\centering
\includegraphics[width=0.48\textwidth]{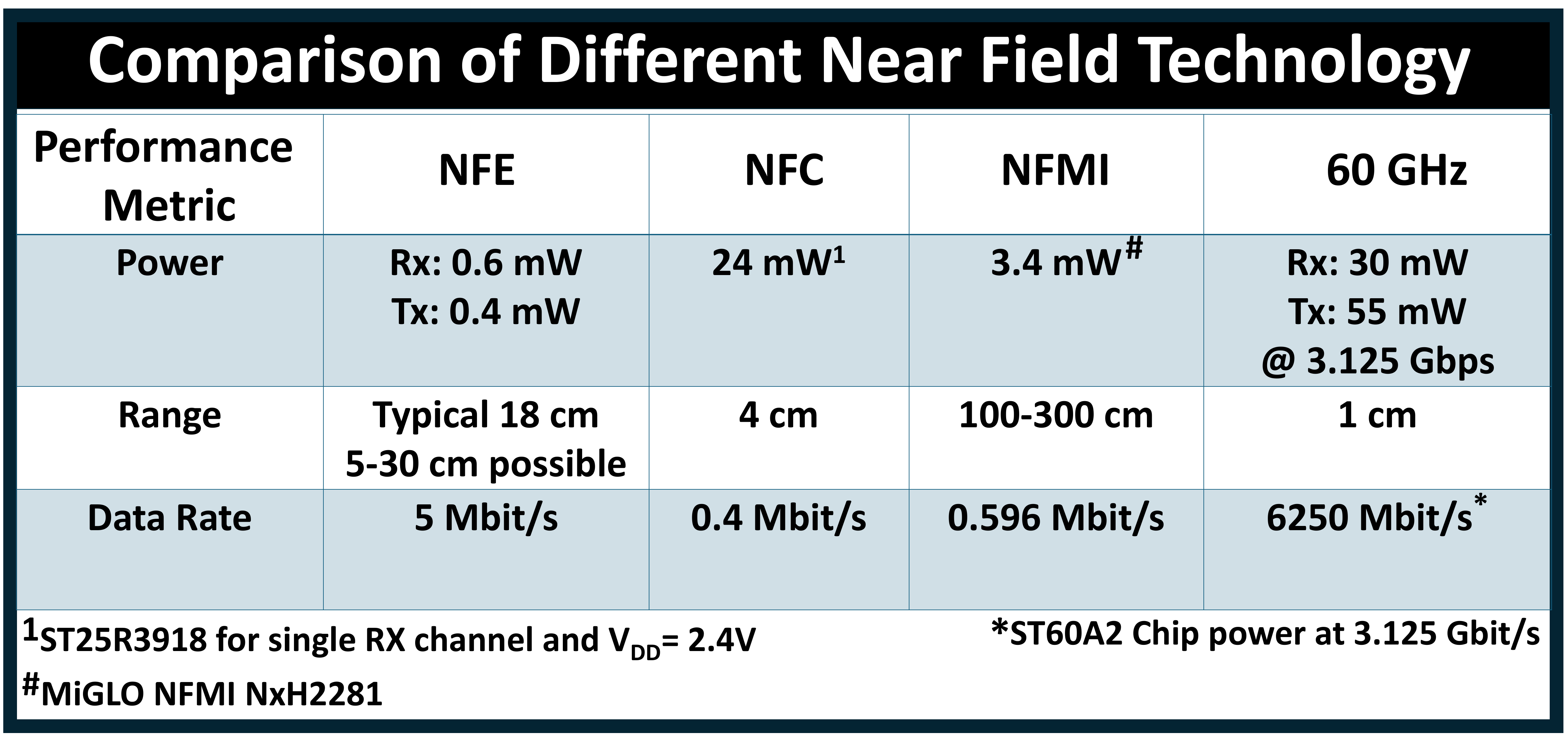}
\caption{Comparison Chart of different NF technologies based on their Power, Range and Data Rate.}
\label{fig:Comparison Chart}
\end{figure}

\section{Comparison of Performance Metrics}
Here we present a comparison of near-field (NF) technologies based on their power consumption, communication range, and data rate. The analysis of power consumption relative to transmission distance for various NF techniques shows that the proposed Wi-R NFE technique offers the lowest power solution. This is illustrated in Figure \ref{fig:Performance Comparison}, which compares its performance with existing technologies such as NFC, NFMI, and mm-Wave (60 GHz) at a selected data rate of 5 Mbps. This makes Wi-R NFE a promising candidate for low-power next-generation wireless networks. Additionally, the results comparing different NF technologies in terms of consumed power, transmission range, and supported data rate are summarized in Figure \ref{fig:Comparison Chart}.

\section{Conclusion}
This work demonstrates near-field electric (NFE)-based wireless communication that facilitates the competing demands of security, low energy, and high throughput for contact-range interactions. The presented NFE-based transceiver establishes a physically localized link using a contained electric field, making proximity the primary security token and removing the need for complex software pairing. Measured power consumption is 0.6 mW for the Rx and 0.4 mW for the Tx, combined is $\sim$24$\times$ lower than NFC and $\sim$3$\times$ lower than NFMI at equivalent data rates. The range is configurable from 5–30 cm, with a typical value of 18 cm. Sustained net throughput reaches $>$3 Mbps at short range. These results show that NFE can support multi‑megabit data rates at $<$1 mW of consumed power at both Tx and Rx, outperforming NFC and NFMI in terms of speed and transmission range. Systematic experiments at different data rates, in both off‑body and on-body scenarios, validate reliable link behavior with low packet error rate and consistently higher throughput. This reliability supports wearable ecosystems, medical devices, touch‑driven exchanges, and human‑centric IoT. The transmission range is directional and data-rate-dependent. An extended coverage up to $>$1 m in the presence of a human body is observed at a selected data rate of 1 Mbps. These characteristics highlight practical tradeoffs between range, data rate, and power that designers can exploit. The NFE approach, therefore, offers a flexible, energy‑efficient solution for short‑range, high‑throughput links. To realize broad adoption, future work should address standardization, interoperability with existing NF technologies, robust multi‑device coordination, and deeper security and regulatory analyses.

\ifCLASSOPTIONcaptionsoff
  \newpage
\fi

\bibliographystyle{IEEEtran}
\bibliography{Near_Field_Electric_Communication}

\end{document}